\documentclass[slac_one]{revtex4}
\usepackage{graphicx}
\usepackage{fancyhdr}
\usepackage{epsfig}
\vspace{1.0cm}

\newcommand{\Frac}[2]{\frac{\displaystyle #1}{\displaystyle #2}} 
\newcommand{\Br}{{\cal B}} 
\newcommand{\cL}{{\cal L}}

\newcommand{\cO}{{\cal O}}

\newcommand{\no}{\nonumber}

\newcommand{\be}{\begin{equation}}
\newcommand{\ee}{\end{equation}}
\newcommand{\bea}{\begin{eqnarray}}
\newcommand{\eea}{\end{eqnarray}}
\newcommand{\beq}{\begin{equation}}
\newcommand{\eeq}{\end{equation}}
\newcommand{\beqa}{\begin{eqnarray}}
\newcommand{\eeqa}{\end{eqnarray}}
\newcommand{\ba}{\begin{array}}
\newcommand{\ea}{\end{array}}

\newcommand{\ccdot}{ \! \cdot \! }

\newcommand{\yuk}{{Y}}

\begin{document}

\title{Status of CP violation in Kaon systems}

\author{Giancarlo D'Ambrosio }
\affiliation{INFN,
Sezione di Napoli, \\
Complesso Universitario di Monte S. Angelo, Via Cintia, Edificio 6,
80126 Naples, Italy }

\begin{abstract}
CP violation is an important tool to test the Standard Model and his extensions.
We describe kaon physics observables testing CP violation and more in general short distance physics. Channels under consideration will be $K \to\pi \nu\bar\nu$, $K \to\pi  l^+ l^- $, $K^\pm\to3\pi$, $K^\pm\to\pi ^\pm \pi \gamma$ and $K^\pm\to\pi ^\pm \pi^0  e^+ e^- $.
\end{abstract}
\maketitle

\thispagestyle{fancy}

\section{INTRODUCTION}

The Standard Model (SM) is very successful phenomenologically; this success has been strengthened  by  Higgs discovery along the potential possibility to have discovered  an ultimate theory up to almost the GUT scale \cite{EliasMiro:2011aa}.
Flavor physics has the possibility to test extensions of the SM in the two possible options, minimal flavor violation (MFV) discussed in the next section or adding new flavor structures. Particularly useful to this purpose are the 
$K^+\to\pi^+\nu\bar\nu$ decays discussed in section III; in section IV we discuss the challenging $K_L \to \pi ^0 e^+ e^-$ and the related channels $K \to \pi  \gamma \gamma$ and others, all interesting as chiral tests too. In section V and VI we  analyze
CP violation and chiral tests in $K^+\to 3\pi$,  $K \rightarrow \pi \pi \gamma $ and  $K \rightarrow \pi \pi e e $ decays.

\section{MINIMAL FLAVOUR VIOLATION}
Flavour physics has been crucial to dismantle arbitrary extensions of the  SM, in fact soon after the discovery of technicolor, an interesting global symmetry, minimal flavour violation  (MFV), was introduced 
to avoid large FCNC carried by the new flavour structures of techni-particles \cite{Chivukula:1987py}.
The SM lagrangian has an interesting symmetry in the limit that all the fermionic sector is massless: defining respectively with $Q$'s, $U$'s and $D$'s, the left-handed doublets,  right-handed up singlets and   right-handed down singlets, the global symmetry ,
${\cal G} _F= U(3)_Q \times U(3)_U \times U(3)_D$, is conserved. This global symmetry is broken by the mass terms

\beq
{\cal L}^\yuk _{SM}\  ~ = ~ \  {\bar Q} {\yuk_D} D  H
+ {\bar Q}  {\yuk_U} U  H_c {{ \rm ~+~h.c.}}~
\eeq
The Higgs generates the mass terms and the Yukawas furnish only sources of flavour group breaking ${\cal G} _F$; then the effective
FCNC   hamiltonian is generated  through this breaking  
\beq {\cal H}_{\Delta F=2}^{SM}\sim {\frac {G_F^2 M _{W}^2}{16\pi^2} }
\left[\frac{{(V_{td}^* m _t ^2 V_{tb})^2}}{v^4}
(\bar{d}_L \gamma ^\mu b_L)^2 + \frac{{(V_{td}^* m _t ^2 V_{ts})^2}}{v^4}
(\bar{d}_L \gamma ^\mu s_L)^2\right]+{\rm charm} \label{eq: FCNC}
\eeq
New flavour structures in susy generated for instance by soft breaking terms
\beq-\cL _{soft}=
 \tilde{Q} ^\dag { m_Q ^2} \tilde{Q}+ \tilde{L}^\dag { m_L^2} \tilde{L}+
\tilde{\bar{U}}{a_u}\tilde{Q}\ H_u+...\eeq
and for generic squark masses, the requirement not to alter  the experimental FCNC status sets a severe limit ($\sim 100 $ TeV) 
to  SUSY scale.  One then requires that   the  New Physics flavour structures  have the same SM flavor breaking, {\it i.e.} the Yukawas, to an effective hamiltonian proportional to eq.(\ref{eq: FCNC}). This effective approach to flavour physics beyond the Standard Model  is the so called minimal flavor violation (MFV) \cite{Chivukula:1987py,Hall:1990ac,Buras:2000dm,MFV}.

\section{THE ULTRA-RARE DECAY $K^+\to\pi^+\nu\bar\nu$}
\unboldmath 

The SM predicts the $V-A\otimes V-A$ effective hamiltonian 
\beqa
{\cal H}& = &\frac{G_{F}}{\sqrt{2}} \frac{\alpha}{2 \pi \sin ^2 \theta_W}
(\ \underbrace{V_{cs}^{*}V_{cd}\ X_{NL}}_{\textstyle{\lambda x_c}} \,
+ \underbrace{ V_{ts}^{*}V_{td}X(x_t)}_{\textstyle{A^2\lambda ^5 \ 
(1-\rho -i{\eta}){ x_t }}}) \ \overline{
s}_L \gamma _\mu d_L \  \overline{\nu } _L \gamma ^\mu \nu _L , 
\label{ampsd} \\ \no
\eeqa
 $x_q=m_q^2/M_W^2$, $\theta_W$ the Weak angle and 
 $X$'s are the Inami-Lin functions with 
Wilson coefficients known at two-loop electroweak corrections \cite{Gorbahn:2011pd}. 
$SU(2)$ isospin symmetry relates hadronic matrix elements for $%
K\rightarrow \pi \nu \overline{\nu }$ to $K\rightarrow \pi l\overline{\nu }$
to a very good precision \cite{Komatsubara:2012pn} while long distance contributions and QCD corrections are under control \cite{Gorbahn:2011pd} and 
the main uncertainties
is due to the strong corrections to the charm loop contribution.
The structure in (\ref{ampsd}) leads to a pure CP violating contribution
to $K_{L}\rightarrow \pi ^{0}\nu \overline{\nu },$ induced only from the top
loop contribution and thus proportional to $\Im m(\lambda _{t} )$
($\lambda _t= V_{ts}^{*}V_{td}$) and free of
hadronic uncertainties. This leads to the prediction \cite{Gorbahn:2011pd}
\begin{equation}
\Br (K^{^{\pm }}\rightarrow \pi ^{\pm }\nu \overline{\nu })_{SM}= (8.22\pm0.69 \pm 0.29) \times 10^{-11} 
\qquad \Br (K_{L}\rightarrow \pi ^{0}\nu \overline{\nu })_{SM} =(2.43^{+0.40} _{-0.37}+ ± 0.06)  \times 10^{-11}
 \label{eq:klpi0nunu}
\end{equation}
where the first is  the parametric uncertainty due to  the error on 
$|V_{cb}|$, $\rho$ and $\eta$, $f_K$,  and the second error summarizes the  theoretical uncertainties on non-perturbative physics and QCD higher order terms.
$K^{^{\pm }}\rightarrow \pi ^{\pm }\nu \overline{\nu }$ receives CP
conserving contributions proportional to $\Re e(\lambda _{c}),$ and to 
$\Re e(\lambda _{t})$ and a CP violating  one proportional to $~~\Im m(\lambda _{t}).$ 
E949 Collaboration \cite{Artamonov:2008qb} and  {\rm E391a } Collaboration \cite{Ahn:2009gb}  have then measured 

\begin{equation}
\Br(K^{^{\pm }}\rightarrow \pi ^{\pm }\nu \overline{\nu })=\left(
1.73_{-1.05}^{+1.15}\right) \times 10^{-10} \qquad {\rm E949}\label{eq:E949}
\end{equation}
\begin{equation}
\Br (K_{L}\rightarrow \pi ^{0}\nu \overline{\nu })  < 2.6  \times 10^{-8} \ {\rm at } \  90 \%  \  {\rm C.L. } \quad {\rm E391a Collaboration}  \end{equation}
The direct   upper bound  for the neutral decay
  can  be improved with a theoretical analysis:
 the isospin structure of any $\overline{s}d$ operator (bilinear in the quark
fields) leads to the model independent relation among 
$A(K_{L}\rightarrow \pi ^{0}\nu \overline{\nu } )$ and 
$A(K^{^{\pm }}\rightarrow \pi ^{\pm }\nu \overline{\nu })$
\cite{Grossman:1997sk}; this leads to   
\beq
\Br(K_{L}\rightarrow \pi ^{0}\nu \overline{\nu })<
4 \ \Br (K^{^{\pm }}\rightarrow \pi ^{\pm }\nu \overline{\nu }) \eeq

The  upcoming KOTO experiment \cite{KOTOweb,Komatsubara:2012pn} for  $K_{L}\rightarrow \pi ^{0}\nu \overline{\nu }$, NA62 \cite{SozziCKM2010,Goudzovski} and possibly ORKA experiment at Fermilab \cite{ORKA} for $(K^{^{\pm }}\rightarrow \pi ^{\pm }\nu \overline{\nu })$  
encourage theoretical investigations of extensions of the SM:  these experiments  probe deeply to the MFV scale \cite{MFV}.
More aggressive NP models can furnish substantial enhancements and be either discovered or ruled out \cite{Gorbahn:2011pd,NPkpnunu}!

 \section{$K_L \to \pi ^0 e^+ e^-$, the related channels $K \to \pi  \gamma \gamma$ and  $K_S \to \pi ^0 e^+ e^-$}
 
The electroweak short distance contribution to $K_L \to \pi ^0 e^+ e^-$, analogously to the one $K_L \to \pi ^0 \nu \bar{\nu}$ is a direct CP violating one, however there is  long distance contamination  due to electromagnetic interactions: i) a CP conserving contribution due to two-photon exchange and ii) an indirect CP violating contribution mediated by one photon exchange, {\rm i.e.}  the contribution suppressed by  $\epsilon $ in $K_L \sim  K_2 + \epsilon K_1 \to \pi ^0 e^+ e^-$ 
determined by the CP conserving $A ( K_S \to \pi ^0 e^+ e^-) $\cite{EPR,DEIP,Buchalla:2003sj}. 

The CP-conserving decays $K^\pm (K_{S}) \rightarrow \pi ^{\pm} 
(\pi ^{0})\ell ^{+}\ell ^{-}$ are dominated by the 
long-distance process $K\to\pi    \gamma ^*  \to \pi   \ell^+ \ell^-$ \cite{EPR,DEIP}.
Our ignorance in the long distance dominated g $A ( K_S \to \pi ^0 l^+ l^-) $ can be parametrized by one parameter
$ a_S$ to be determined experimentally, NA48,  finds respectively in the electron \cite{Batley:2003mu} and muon final state  \cite{Batley:2004wg} 
 \begin{equation}
  | a_S|_{e e } = 1.06 ^{+0.26} _{-0.21} \pm 0.07 
 \qquad
 |a_S|_{\mu \mu } = 1.54 ^{+0.40} _{-0.32} \pm 0.06
\end{equation}
These results  allow us to evaluate the CP violating branching 
\begin{eqnarray}
B(K_{L}\rightarrow \pi ^{0}e^{+}e^{-})_{CPV}\,=\     \left[
15.3\,a_{S}^{2}\,-\,6.8\frac{\displaystyle \Im \lambda _{t}}{\displaystyle %
10^{-4}}\,a_{S}\,+\,2.8\left( \frac{\displaystyle \Im \lambda _{t}}{%
\displaystyle 10^{-4}}\right) ^{2}\right] \times 10^{-12}~,
\label{eq:cpvtot}
\end{eqnarray}
The first term and last terms are respectively the  indirect and the direct contribution, the second one is the interference, expected 
constructive allowing a stronger  signal \cite{Buchalla:2003sj}.

This  prediction is not far from the 
the present bound 
from KTeV \cite{AlaviHarati:2003mr} 
\begin{equation}
\Br(K_{L}\rightarrow \pi ^{0}e^{+}e^{-})<2.8\times 10^{-10}\quad
{\rm at} \quad 90\% \quad {\rm CL}.
\end{equation}
which also sets the interesting limit
$\Br(K_{L}\rightarrow \pi ^{0}\mu ^{+}\mu ^{-})<3.8\times 10^{-10}$ \cite{AlaviHarati:2000hs}.
Still we have to show that we have under control the CP conserving contribution generated by two photon exchange. 
\hspace*{0.1cm} The general amplitude for $K_{L}(p)\rightarrow \pi
^{0}\gamma (q_{1})\gamma (q_{2})$ can be written in terms of two 
Lorentz and gauge invariant amplitudes $A(z,y)$ and $B(z,y):$
\begin{eqnarray}
  && {\cal A}( K_{L}\rightarrow \pi ^{0}\gamma \gamma ) =  
   \frac{G_{8} \alpha }{ 4\pi }
   \epsilon_{1 \mu} \epsilon_{2 \nu} 
   \Big[  A(z,y) 
    (q_{2}^{\mu}q_{1}^{\nu }-q_{1}\ccdot q_{2}~ g^{\mu \nu } ) +  \nonumber \\
 && \qquad + 
   \frac{2B(z,y)}{m_{K}^{2}} 
    (p \ccdot q_{1}~ q_{2}^{\mu }p^{\nu} + p\ccdot q_{2} ~p^{\mu }q_{1}^{\nu}
   -p\ccdot q_{1}~ p\ccdot q_{2}~ g^{\mu \nu } 
   - q_{1}\ccdot q_{2}~ p^{\mu }p^{\nu } ) \Big]~, \quad
\label{eq:kpgg}
\end{eqnarray}
where $y=p (q_{1}-q_{2})/m_{K}^{2}$ and $z\,=%
\,(q_{1}+q_{2})^{2}/m_{K}^{2}$.
Then the double differential rate is given by 
\begin{equation}
\frac{\displaystyle \partial ^{2}\Gamma }{\displaystyle \partial y\,\partial %
z}\sim [%
\,z^{2}\,|\,A\,+\,B\,|^{2}\,+\,\left( y^{2}-\frac{\displaystyle \lambda
(1,r_{\pi }^{2},z)}{\displaystyle 4}\right) ^{2}\,|\,B\,|^{2}\,]~,
\label{eq:doudif}
\end{equation}
where $\lambda (a,b,c)$ is the usual kinematical function 
and $r_{\pi }=m_{%
\pi }/m_{K}$.
 Thus in the region of small $z$ (collinear photons) the $B$
amplitude is dominant and can be determined separately from the $A$
amplitude. This feature is crucial in order to disentangle the CP-conserving
contribution $K_{L}\rightarrow \pi ^{0}e^{+}e^{-}$. In fact  
the lepton pair 
produced by  photons  in $S$-wave, like 
an ${ A}(z)$-amplitude, are suppressed  by the lepton mass while
the photons  in $B(z,y)$ are
also in $D$-wave and so the resulting  $K_{L}\rightarrow \pi ^{0}e^{+}e^{-}$
 amplitude, 
$A(K_{L}\rightarrow \pi ^{0}e^{+}e^{-})_{CPC}$,  
 does not suffer from  the electron mass suppression \cite{Buchalla:2003sj}.
The important message is that experiments by  studying  the $K_{L}\rightarrow \pi
^{0}\gamma \gamma  $  $z-$spectrum  have been able to limit  
$\Br(K_L\rightarrow\pi^0 e^+ e^-)<5\cdot 10^{-13}$ {\rm at} \quad 90\% \quad {\rm CL}
 \cite{Lai:2002kf,Abouzaid:2008xm}.

\begin{figure}[htb]
\vfill
\begin{minipage}[b]{7cm}\centering
\mbox{\epsfig{file=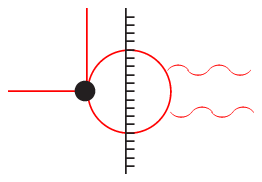,width=6cm,height=5cm}} \vspace{0.4cm}
\caption{ Unitarity contributions to $K \rightarrow \pi  \gamma \gamma $} 
\label{fig:BKpiggbis}
\end{minipage}
\hspace{0.8cm}
\begin{minipage}[b]{6.5cm}\centering
\mbox{\epsfig{file=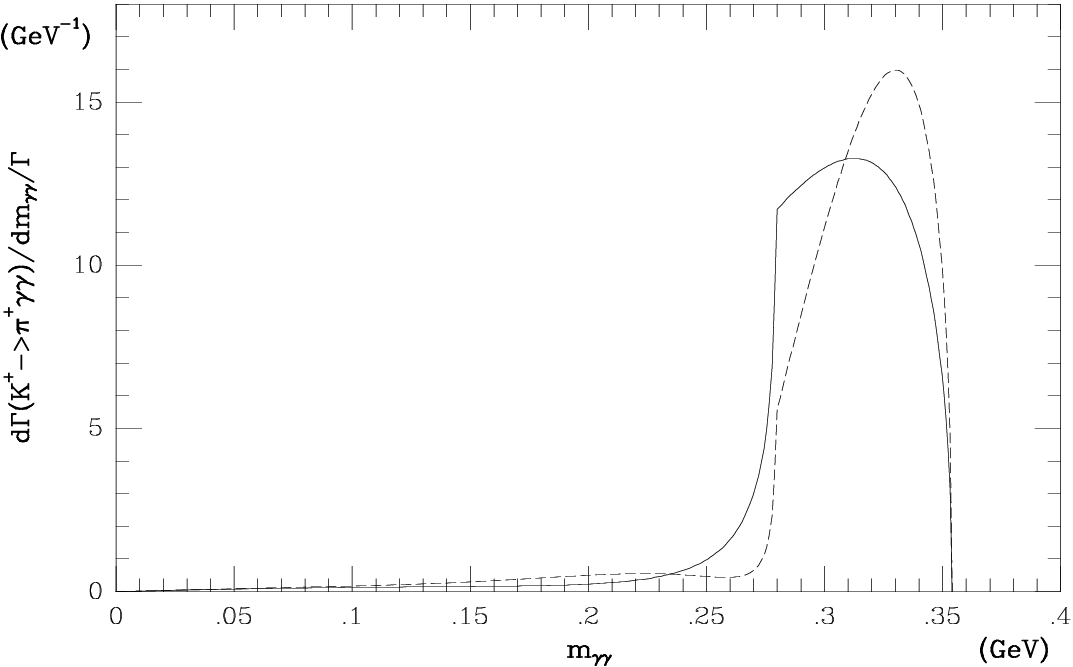,width=6.5cm,height=5.5cm}}
\caption{$K^{+}\rightarrow \pi ^+ \gamma \gamma $: $\hat{c}=0$ , full   line, $\hat{c}=-2.3$ , dashed   line,    \cite{D'Ambrosio:1996zx}} 
\label{fig:plotbis}
\end{minipage}
\vfill
\end{figure}

Recently a related channel, $K^{+}\rightarrow \pi ^{+}
\gamma \gamma $,  has attracted attention: 
new  measurements of this decay have been performed using minimum bias data sets collected during a 3-day special NA48/2 run in 2004 with 60 ${\rm GeV}$  $K^{\pm}$  beams, and a 3-month NA62 run in 2007 with 74 GeV/c $K^{\pm}$ beams \cite{Goudzovski}.

 This channels start at  $\mathcal{O(}p^{4})$,   with     pion (and kaon) loops    and a local term $\hat{c}$. 
 Due to the presence of the pion pole, there is a new helicity amplitude, $C$ \cite{Ecker:1987hd};  the unitarity contributions  at  $\mathcal{O(}p^{6})$ 
in Fig.1 enhance the amplitude  $A$ by 30\%-40\% , along with the generation of  $B$-type amplitude \cite{D'Ambrosio:1996zx}; the differential decay rate is 
\noindent
\begin{equation}\Frac{d^2\Gamma}{dy dz}\sim \left[  {z^2}({|}A+ {B}{|}^2 +|{C}|^2 ) {+}\left(
y^2-\left(\Frac{(1+r_\pi^2-z)^2}{4}-r_\pi^2\right)\right)^2{|}{B} {|^2}\right]   \end{equation}

The constant $\hat{c}$ can be fixed by a  precise determination of the rate and the spectrum as shown in Fig.2 \cite{D'Ambrosio:1996zx}; this constant, combination of strong and weak counterterm, is predicted to have contributions from the axial spin-1 contributions. 

$$\hat{c}=\Frac{128 \pi ^2 }{3}\left[3(L_9 +L_{10}) +N_{14}-N_{15}-2N_{18})  \right]\stackrel{\rm FM  }{=}2.3\  (1-2\ k_f)\ ,$$
with $k_f$ is the factorization factor in the FM model   or the weak  axial  vector  coupling of Ref. \cite{DP98}.   BNL 787  got 31 events leading to  $B(K^+ \to \pi ^+ \gamma \gamma)  \sim (6\pm1.6)\cdot 10^{-7}$ \cite{Kitching:1997zj}
and a value of $\hat{c}=1.8\pm0.6$.
Recently NA48 has presented   preliminary results normalizing $K^+ \to \pi ^+ \gamma \gamma$ with the channel $K^+\rightarrow\pi^+  \pi^0$:  $\Br(K^+ \to \pi ^+ \gamma \gamma) =(1.01\pm0.04\pm 0.06)\cdot 10^{-6}$  and $\hat{c}=2.00 \pm 0.24_{stat} \pm 0.09_{syst}$  \cite{Goudzovski}.

 \section{CP aymmetries in $K^+\to 3\pi$-decays  }
Direct CP violation in charged kaons is subject of extensive researches at NA48/2 \cite{SozziCKM2010}. 
Studying 
the $K\rightarrow 3 \pi$ Dalitz distribution in  $Y,X$ \cite{DI98,Beringer:1900zz}
$$
|A(K\rightarrow 3 \pi)|^2 \sim 1+g\ Y +j\ X + {\cal O}(X^2, Y^2)
$$
and determining  both  charged kaon  slopes, $g_{\pm} $, we can define the slope charge asymmetry: 
\beq
\Delta g/2g=(g_+-g_-)/(g_+ + g_-) \label{eq:deltag}.\eeq
There are two independent $I=1$ isospin amplitudes $(a, b)$, 
\beq
A(K^{+} \to 3\pi) = {a} e^{i \alpha_0} + {b} e^{i \beta_0 } Y
+ \cO(Y^2,X^2) 
\label{eq:dg} 
\eeq
with corresponding
final state interaction phases, $\alpha_0$ and $\beta_0$.
 The hope is that $\Delta g$
in  (\ref{eq:deltag}) does {\bf NOT} need to be suppressed by a $\Delta I = 3/2$ transition.
The strong
 phases, generated by the $2 \to 2$ rescattering, 
 actually have their own
kinematical dependence \cite{IMP} and can be expressed in terms 
of  the Weinberg scattering lenghts, $a_0$ and  $a_2$.
It is particularly interesting to estimate the  Standard Model (SM) size
 for $\Delta g/{2g}$,  valid if there is a good chiral expansion 
for the CP conserving/violating $a,b$ amplitudes \cite{DI98,IMP,DIP}:
\beqa
\frac{\Delta g}{2g}\sim 22 \epsilon '(\alpha_0 - \beta_0)\sim 10^{-5}.\nonumber 
\eeqa
 The $K^+\rightarrow \pi ^{+} \pi ^{0} \pi ^{0}$ 
NA48/2 resut \cite{Batley:2006mu} and  New Physics (NP) scenarios
 \cite{D'Ambrosio:1999jh}
\beqa
\frac{\Delta g}{2g}
\stackrel{{\rm NA48/2}}{=}
(1.8\pm2.6)\cdot 10^{-5}\qquad\stackrel{{\rm NP}}{\le}
10^{-4}.\nonumber 
\eeqa
can then be compared to the SM.

\begin{figure}[htb]
\vfill
\begin{minipage}[b]{7cm}\centering
\mbox{\epsfig{file=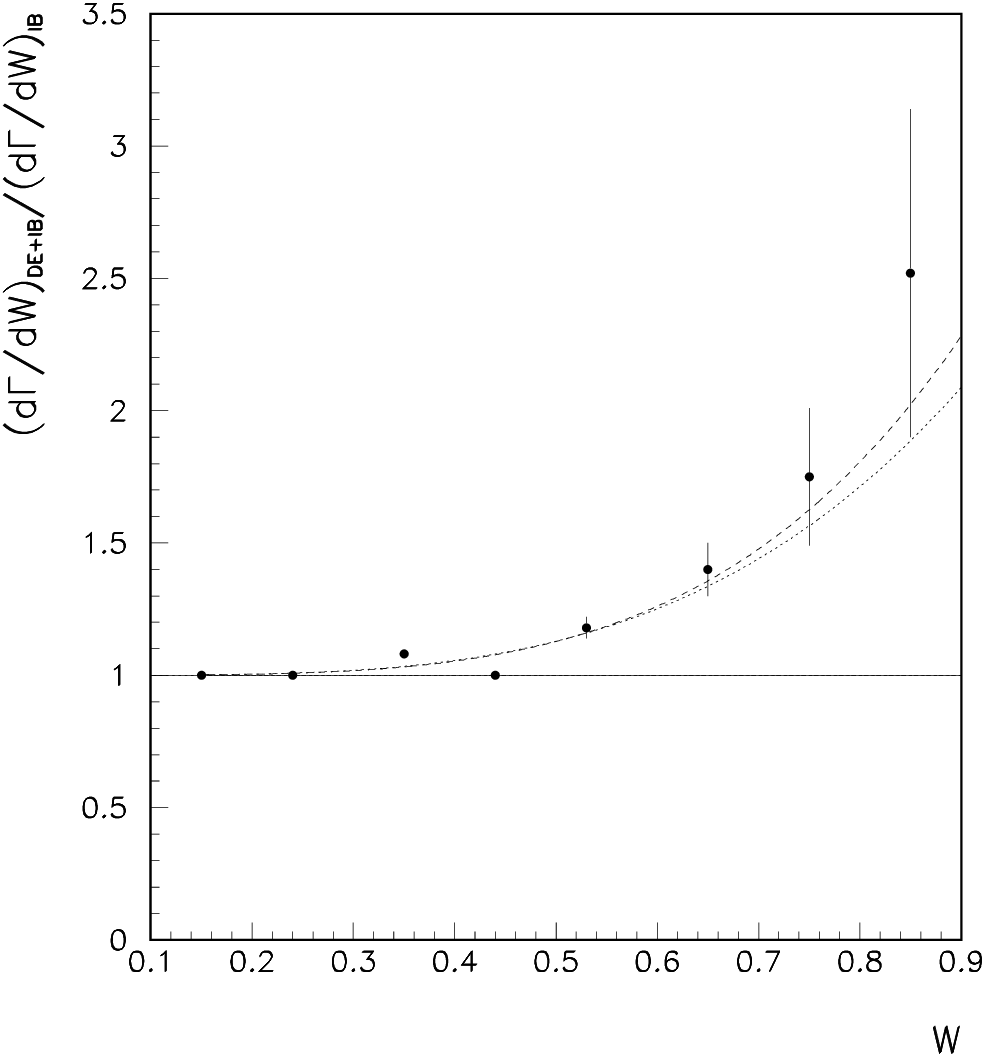,width=6cm,height=5cm}} \vspace{0.4cm}
\caption{deviations from IB } 
\label{fig:IBDE}
\end{minipage}
\hspace{0.8cm}
\begin{minipage}[b]{6.5cm}\centering
\mbox{\epsfig{file=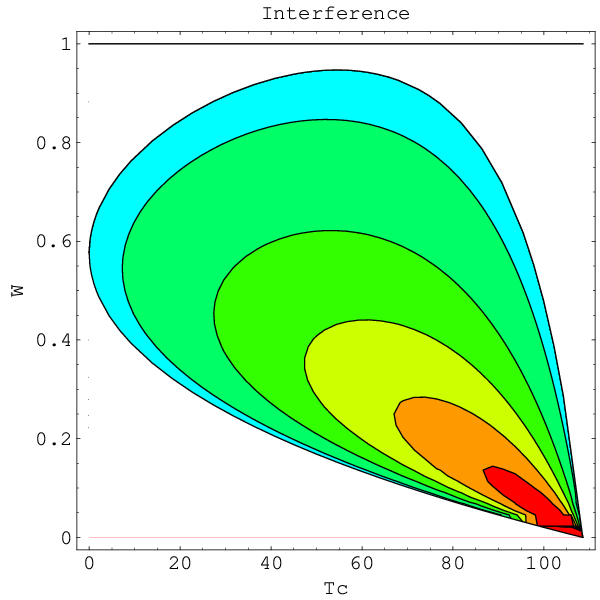,width=6.5cm,height=5.5cm}}
\caption{ $T_c^*-W$-Dalitz plot. In this contour plot of the interference Branching the red area corresponds to more dense and thus larger
contribution}
\label{fig:bo}
\end{minipage}
\vfill
\end{figure}

\section{   $K \rightarrow \pi \pi \gamma $ and  $K \rightarrow \pi \pi e e $-decays}
CP violation has been also studied in the $K \rightarrow \pi \pi \gamma $ and  $K \rightarrow \pi \pi e e $ decays.
We can decompose
$K(p)\rightarrow \pi (p_{1})\pi (p_{2})\gamma (q)$  decays,
according to gauge and Lorentz invariance,
in electric ($E)$ and magnetic ($M)$ terms \cite{Cappiello:2007rs}. 
In the electric transitions one generally separates the bremsstrahlung
amplitude $E_{B}$, predicted   by the Low theorem  in 
terms of the non-radiative amplitude and enhanced by the $1/E_\gamma$ 
behavior.
Summing over photon
helicities: 
${{\mbox{d} ^2\Gamma }/({\mbox{d}z_{1}\mbox{d}z_{2}}}) \sim 
|E(z_{i})|^{2}+|M(z_{i})|^{2}$. At the lowest order, ($p^{2})$,  one
obtains only $E_{B}$.
Magnetic and electric direct emission amplitudes
 can be decomposed in a multipole expansion. In  Table 2
  we show the present  experimental status of the DE amplitudes
and the leading multipoles.

\centerline{\bf Table 2  $ {DE_{exp}} $}
\[
\begin{array}{ccc} 
K_S\rightarrow\pi^+\pi^-\gamma & <9\cdot 10^{-5}&
{E1} \\ 
K^+\rightarrow\pi^+\pi^0\gamma&
(0.44\pm0.07)10^{-5}
& M1,{E1} \\ 
K_L\rightarrow\pi^+\pi^-\gamma & 
(2.92\pm0.07)10^{-5}
&{M1,} {\rm VMD}
\end{array}
\] 
 Particularly interesting   are the recent interesting NA48/2 data regarding
  $K^{+}\rightarrow \pi ^{+}\pi ^{0}\gamma $ decays  \cite{Batley:2010aa}.
Due to the $\Delta I=3/2$ suppression of the bremss\-trahlung,   
in\-ter\-fer\-ence between $E_B$ and $E1$  and magnetic tran\-si\-tions
can be measured. Defining  $z_{i}={p_{i}\cdot q}/m_{K}^{2} \quad z_{3}=p_{_{K}}\cdot q /{m_{K}^{2}}$
 and
$z_{3}z_{+}=\frac{m_{\pi ^{+}}^{2}}{m_{K}^{2}}W^{2}$
we can 
study  the deviation from bremsstrahlung from the decay distribution  
\begin{eqnarray}
\Frac{\partial ^{2}\Gamma }{\partial T_{c}^{\ast }\partial W^{2}}
&=\Frac{\partial ^{2}\Gamma _{IB}}{\partial T_{c}^{\ast }\partial
W^{2}}\left[1+\Frac{m_{\pi ^{+}}^{2}}{m_{K}}2Re\left(\Frac{E_{DE}}{eA}\right)W^{2}
 +\Frac{m_{\pi^{+}}^{4}}{m_{K}^{2}}\left(\left|\Frac{E_{DE}}{eA}\right|^{2}+\left|\Frac{M_{DE}}{
eA}\right|^{2}\right)W^{4}\right], \nonumber
\end{eqnarray}
where $A=A(K^{+}\rightarrow \pi ^{+}\pi ^{0})$;  we plot in Fig. 3 this experimental   deviation from  bremsstrahlung. The Dalitz plot  distribution of the interference term is shown  in  Fig. 4. Study of the Dalitz plot has lead 
NA48 to these results  \cite{Batley:2010aa} 
\centerline{\bf Table 3}
\begin{center}
{\hskip0,8cm NA48/2} \hskip1.0cm $T_{c}^{\ast }\in \left[0, 80\right]$\ MeV\\ 
\begin{tabular}{ccc}
\hline 
{\it Frac(DE)}&=&$(3.32{{\pm} }0.15\pm 0.14){{\times} }10^{-2}$\\
{\it Frac(INT)}&=&$(-2.35{{\pm} }0.35\pm 0.39){{\times} }10^{-2}$\\
\end{tabular}
\end{center}
Also the  interesting CP bound was obtained \cite{Batley:2010aa}:  
\beq\Frac{\Gamma(K^{+}\rightarrow \pi ^{+}\pi ^{0}\gamma )- \Gamma(K^{-}\rightarrow \pi ^{-}\pi ^{0}\gamma )}{\Gamma(K^{+}\rightarrow \pi ^{+}\pi ^{0}\gamma )+\Gamma(K^{-}\rightarrow \pi ^{-}\pi ^{0}\gamma )}<1.5\cdot 
10^{-3} \quad{\rm at} \quad 90\% \quad {\rm CL}.\eeq
With more statistics the Dalitz plot analysis in Fig. 4 will be more efficient.

 \begin{figure}[t]
\begin{center}
\includegraphics[width=5.5cm]{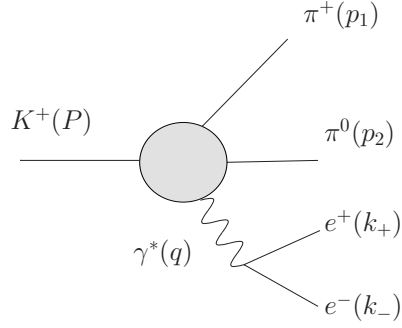}
\end{center}
\caption{\small{\it{Photon-mediated $K^+\to\pi^+\pi^0e^+e^-$ decay with our kinematical conventions. The blob represents the hadronic tensor $H_{\mu}$.}}}
\end{figure} 

 We have studied also the decay $K^\pm \to \pi^\pm \pi^0 e^+e^-$ in Fig. 5 \cite{Cappiello:2011qc}.
 Historically kaon four body semileptonic decays, $K_{e4}$ have been studied as a tool to tackle final state rescattering effects in $K\to \pi \pi$-decays: crucial to this goal has been  finding an appropriate set of kinematical variables  which would allow i) to treat   the system as two body decay in dipion mass $M_{\pi \pi}$ and dilepton mass $M_{l^+ l^-}$ \cite{Cabibbo:1965zz} and ii) to identify appropriate kinematical asymmetries to extract observables crucially dependent on final state interaction. 
 \begin{figure}[htb]
\vfill
\begin{minipage}[b]{8cm}\centering
\mbox{\epsfig{file=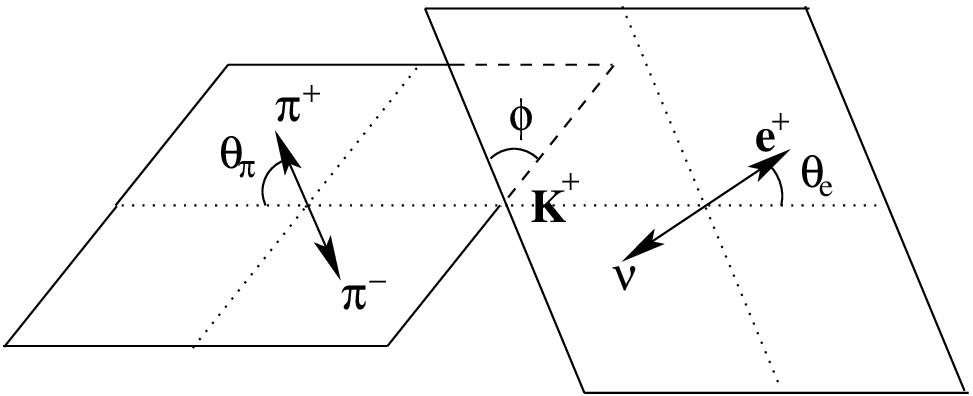,width=8.5cm,height=5cm}} \vspace{0.4cm}
\caption{ $K^+ \to   \pi ^+ \pi^0 e ^+  e^-$  kinematical planes:  N.~Cabibbo and A.~Maksymowicz defintion of the angles \cite{Cabibbo:1965zz}} 
\label{fig:diplane}
\end{minipage}
\hspace{0.8cm}
\begin{minipage}[b]{8.5cm}\centering
\mbox{\epsfig{file=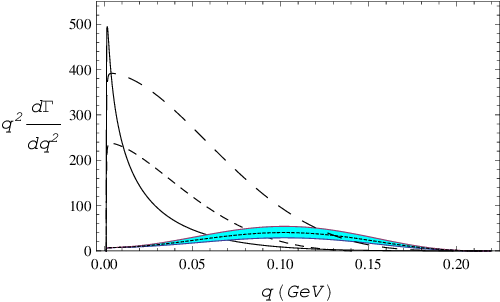,width=8.5cm,height=5.5cm}}
\caption{$q$ dependence of the different contributions. The solid line represents the Bremsstrahlung. The dashed lines (from bigger dash to smaller dash) are $100\times$M, $100\times$BE and $300\times$E, respectively. } 
\label{fig:plotter}
\end{minipage}
\vfill
\end{figure} 
In Fig. \ref{fig:diplane} we show the traditional kinematical variables for the four body kaon semileptonic decay which allow to write the four body phase space $\Phi$ in terms of the two two-body phase space $\Phi_{\pi}$ $\Phi_{\ell}$  from \cite{Cabibbo:1965zz}

\begin{eqnarray}
d\Phi &=\frac{1}{4m_K^2}(2\pi)^5\int d s_\pi \int d
s_\ell \lambda^{1/2}(m_K^2,p_{\pi}^2,q^2)\Phi_{\pi}\Phi_{\ell}. 
\end{eqnarray}  

Then defining $q^2=M_{e\nu}^2$ and $p_{\pi}^2$ the $\pi\pi$ invariant  mass
 we can write
\begin{equation}\label{ps31}
d^5\Phi=\frac{1}{2^{14}\pi^6m_K^2}\frac{1}{s_\pi}\sqrt{1-\frac{4m_\ell^2}{q^2}}
\lambda^{1/2}(m_K^2,p_{\pi}^2,q^2)\lambda^{1/2}(p_{\pi}^2,m_{\pi^+}^2,m_{\pi^0}^2)d
p_{\pi}^2 d q^2 d\cos\theta_\pi d\cos\theta_\ell d\phi,
\end{equation}

Then the $K_{e4}$
 amplitude is written  as 
\begin{eqnarray}\label{amplitudeCM}
{\cal M}_{l4}=\frac{G_F}{\sqrt{2}} V_{us} \big[\bar{u}(p_e)\gamma^\mu (1-\gamma^5)v(p_\nu)\big] H_{\mu}(p_1,p_2,q),
\end{eqnarray} 
where $H_{\mu}$ is the hadronic vector, which can be written in terms of 3 form factors $F_{1,2,3}$:
\begin{equation}\label{param}
H^{\mu}(p_1,p_2,q)=F_1 p^\mu_1+F_2 p_2^\mu+F_3 \varepsilon^{\mu\nu\alpha\beta}p_{1\nu}p_{2\alpha}q_\beta. \label{FF}
\end{equation}

The goal was to obtain some asymmetry strongly dependent on  the final state
$\delta ^i _j(s)$ in the 
form factors $$F_i (s)=
f_i(s)e^{i{\delta ^0 _0(s)}}+ ..$$
Indeed  
\begin{eqnarray}\label{angular}
\frac{d^5\Gamma}{dE_{\gamma}^*dT_c^*dq^2d\cos\theta_{\ell} d\phi}&={\cal{A}}_1+{\cal{A}}_2\sin^2\theta_{\ell}+{\cal{A}}_3\sin^2\theta_{\ell}\cos^2\phi\nonumber\\
&\!\!\!\!\!\!\!\!\!\!\!\!\!\!\!\!\!\!\!\!\!\!\!\!\!\!\!\!\!\!\!\!\!\!\!\!\!\!\!\!\!\!+{\cal{A}}_4\sin2\theta_{\ell}\cos\phi+{\cal{A}}_5 \sin\theta_{\ell}\cos\phi+{\cal{A}}_6 \cos\theta_{\ell}\nonumber\\
&\!\!\!\!\!\!\!\!\!\!\!\!\!\!\!\!\!\!\!\!\!\!\!\!\!\!\!\!\!\!\!\!\!\!\!\!\!\!\!\!\!\!+{\cal{A}}_7 \sin\theta_{\ell}\sin\phi+{\cal{A}}_8\sin 2\theta_{\ell}\sin\phi+{\cal{A}}_9\sin^2\theta_{\ell}\sin 2\phi,
\end{eqnarray}
where $\theta_{\ell}$ and $\phi$ are two   variables for $K_{l4}$ decays~\cite{Cabibbo:1965zz} and ${\cal{A}}_i$ are dynamical functions that can be parameterized in terms of 3 form factor. ${\cal{A}}_{8,9}$, odd in $\theta_{\ell}$ are also linearly dependent on the final state, establishing a clear way to determine them; while ${\cal{A}}_{5,6,7}$ are generated by interference with the axial leptonic current.

One can easily show that the Bremsstrahlung, direct emission and electric interference terms contribute to ${\cal{A}}_{1-4}$. In contrast, ${\cal{A}}_{8,9}$ receive contributions from the electric-magnetic interference terms (BM and EM) and therefore capture long-distance induced P-violating terms. ${\cal{A}}_{5,6,7}$ are also P-violating terms but generated through the interference of $Q_{7A}$ with long distances.

Essentially two groups  \cite{Sehgal:1992wm} applied the $K_{l4}$ decays to the decay  $K_L \to\pi^+\pi^-e^+e^-$, here the targets are mainly short distance physics, {\it  i.e.}  ${\cal{A}}_{5,6,7}$ 
and the diplane angular asymmetry proportional to ${\cal{A}}_{8,9}$. This last observable is large and has been measured
 by KTeV and NA48  
\cite{KLpipiee_exp,Beringer:1900zz}; however this observable is proportional to 
electric  (bremsstrahlung) and magnetic interference, both contributions known already from $K_L \to\pi^+\pi^- \gamma$; in fact these known contributions are large and they may obscure smaller but more interesting short distance physics effects. 

We have performed a similar analysis for the decay  $K^+\to\pi^+\pi^0e^+e^-$
trying to focus on i) short distance physics and ii) all possible 
Dalitz plot analyses to disentangle 
  all possible interesting long and short distance effects \cite{Cappiello:2011qc}. 
This decay has not been observed yet, and the interesting physics is hidden by bremsstrahlung  \cite{Cappiello:2011qc,Pichl:2000ab}  
\begin{eqnarray}\label{ratioPi} 
\Br(K^+ \to   \pi ^+ \pi^0 e ^+  e^-)_{B}&\sim (330\pm 15)\cdot 10^{-8}\nonumber\\
\Br(K^+ \to\pi ^+ \pi^0 e ^+  e^-)_M&\sim ( 6.14\pm 1.30)\cdot 10^{-8},
\end{eqnarray}
and so Dalitz plot analysis is necessary in order to capture the more interesting direct emission   contributions.
The  $K^+ \to   \pi ^+ \pi^0 e ^+  e^-$-amplitude is written as
\begin{eqnarray}\label{amplitude1}
{\cal M}_{LD}=\frac{e}{q^2} \big[\bar{u}(k_-)\gamma^\mu v(k_+)\big] H_{\mu}(p_1,p_2,q),
\end{eqnarray} 

\begin{figure}[htb]
\vfill
\begin{minipage}[b]{7cm}\centering
\mbox{\epsfig{file=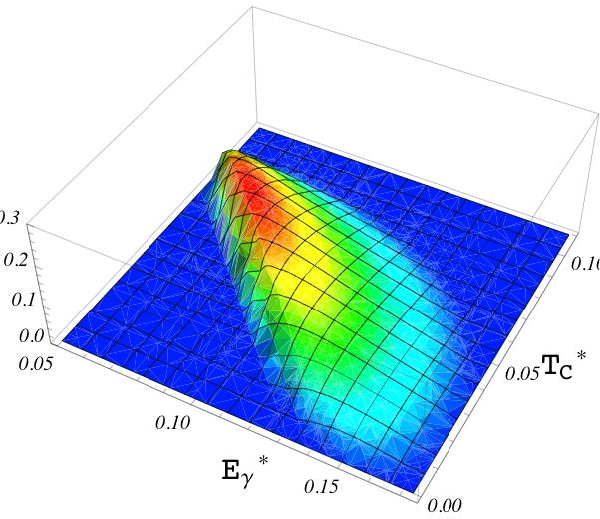,width=6cm,height=5cm}} \vspace{0.4cm}
\caption{ Dalitz plot in the $(E_{\gamma}^{\ast},T_c^{\ast})$ plane at $q^2=(50$ MeV$)^2$ for the P-violating BM contribution} 
\label{fig:BKpigg}
\end{minipage}
\hspace{0.8cm}
\begin{minipage}[b]{6.5cm}\centering
\mbox{\epsfig{file=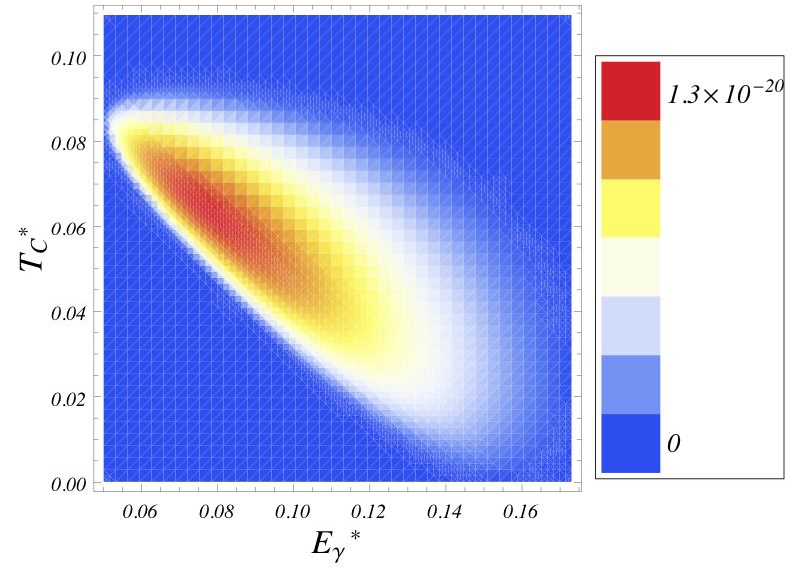,width=6.5cm,height=5.5cm}}
\caption{Dalitz plot BM contribution: two-dimensional density projection} 
\label{fig:plot}
\end{minipage}
\vfill
\end{figure}

We may wonder also what it is the advantage to study this 4-body decay,  $K^+\to\pi^+\pi^0e^+e^-$, versus $K^+ \to   \pi ^+ \pi^0  \gamma$;  in fact there 
are two reasons to investigate this channel, i) first trivially there are more short distance operators and also more long distance observables (for instance interfering electric and magnetic amplitudes) and ii) going to large dilepton invariant mass there is an extra tool compared to $K^+ \to   \pi ^+ \pi^0  \gamma$ to separate the bremsstrahlung component  \cite{Cappiello:2011qc}. For instance at large  dilepton invariant mass the bremsstrahlung can be even 100 time smaller than the magnetic contribution. 
In our paper we give practically all the distributions in eq. (\ref{angular}), here as example we show 
 in Figs. 8 and 9 the Dalitz plot distribution for the novel electric magnetic  interference.
This decay has been analyzed by NA48/2-NA62.

\section{Conclusions}
We are looking forward to the upcoming $K_{L} \to \pi^{0} \nu \bar{\nu}$ KOTO \cite{KOTOweb} and $K^{+} \to \pi^{+} \nu \bar{\nu}$  \cite{Goudzovski} NA62 experiments probing deeply the flavour structure of the SM and we hope ORKA
will join this enterprise \cite{ORKA}.
We have also shown that there are other decay modes like $K_L\rightarrow\pi^0 e^+ e^-$,  $K^{+}\rightarrow \pi ^{+}
\gamma \gamma $
and $K^+\to\pi^+\pi^0e^+e^-$ which are very useful, in particular these last two have been studied recently by NA62.
I would like also to mention  CPT tests in kaon decays \cite{Ambrosino:2006ek}
through Bell-Steinberger relations, recently updated in \cite{Beringer:1900zz}; these leads to best CPT limit and an accurate determination of the CP violating parameter $\epsilon$. 

\section*{Acknowledgements}
I thanks the organizers of the FPCP 2012 workshop for their kindness.  


\end{document}